\documentclass[12pt]{article}
\usepackage{epsfig,epsf}
\renewcommand{\thefootnote}{\fnsymbol{footnote}}
\begin{document}
\def\beq{\begin{equation}}
\def\eeq{\end{equation}}
\def\eq#1{{Eq.~(\ref{#1})}}
\def\fig#1{{Fig.~\ref{#1}}}
\newcommand{\as}{\alpha_S}
\newcommand{\bra}[1]{\langle #1 |}
\newcommand{\ket}[1]{|#1\rangle}
\newcommand{\bracket}[2]{\langle #1|#2\rangle}
\newcommand{\intp}[1]{\int \frac{d^4 #1}{(2\pi)^4}}
\newcommand{\mn}{{\mu\nu}}
\newcommand{\tr}{{\rm tr}}
\newcommand{\Tr}{{\rm Tr}}
\newcommand{\T} {\mbox{T}}
\newcommand{\braket}[2]{\langle #1|#2\rangle}
\newcommand{\ab}{\bar{\alpha}_S}

\setcounter{secnumdepth}{7}
\setcounter{tocdepth}{7}
\parskip=\itemsep               

\setlength{\itemsep}{0pt}       
\setlength{\partopsep}{0pt}     
\setlength{\topsep}{0pt}        
\setlength{\textheight}{22cm}
\setlength{\textwidth}{174mm}
\setlength{\topmargin}{-1.5cm}

%
\renewcommand{\thefootnote}{\fnsymbol{footnote}}
\newcommand{\beqar}[1]{\begin{eqnarray}\label{#1}}
\newcommand{\eeqar}{\end{eqnarray}}
\newcommand{\m}{\marginpar{*}}
\newcommand{\lash}[1]{\not\! #1 \,}
\newcommand{\nn}{\nonumber}
\newcommand{\D}{\partial}
\newcommand{\h}{\frac{1}{2}}
\newcommand{\g}{{\rm g}}
\newcommand{\el}{{\cal L}}
\newcommand{\A}{{\cal A}}
\newcommand{\Ka}{{\cal K}}
\newcommand{\al}{\alpha}
\newcommand{\be}{\beta}
\newcommand{\ep}{\varepsilon}
\newcommand{\ga}{\gamma}
\newcommand{\de}{\delta}
\newcommand{\De}{\Delta}
\newcommand{\et}{\eta}
\newcommand{\ka}{\vec{\kappa}}
\newcommand{\la}{\lambda}
\newcommand{\ph}{\varphi}
\newcommand{\si}{\sigma}
\newcommand{\ro}{\varrho}
\newcommand{\Ga}{\Gamma} 
\newcommand{\om}{\omega}
\newcommand{\La}{\Lambda}  
\newcommand{\tG}{\tilde{G}}
\renewcommand{\theequation}{\thesection.\arabic{equation}}

%
\def\ap#1#2#3{     {\it Ann. Phys. (NY) }{\bf #1} (19#2) #3}
\def\arnps#1#2#3{  {\it Ann. Rev. Nucl. Part. Sci. }{\bf #1} (19#2) #3}
\def\npb#1#2#3{    {\it Nucl. Phys. }{\bf B#1} (19#2) #3}
\def\plb#1#2#3{    {\it Phys. Lett. }{\bf B#1} (19#2) #3}
\def\prd#1#2#3{    {\it Phys. Rev. }{\bf D#1} (19#2) #3}
\def\prep#1#2#3{   {\it Phys. Rep. }{\bf #1} (19#2) #3}
\def\prl#1#2#3{    {\it Phys. Rev. Lett. }{\bf #1} (19#2) #3}
\def\ptp#1#2#3{    {\it Prog. Theor. Phys. }{\bf #1} (19#2) #3}
\def\rmp#1#2#3{    {\it Rev. Mod. Phys. }{\bf #1} (19#2) #3}
\def\zpc#1#2#3{    {\it Z. Phys. }{\bf C#1} (19#2) #3}
\def\mpla#1#2#3{   {\it Mod. Phys. Lett. }{\bf A#1} (19#2) #3}
\def\nc#1#2#3{     {\it Nuovo Cim. }{\bf #1} (19#2) #3}
\def\yf#1#2#3{     {\it Yad. Fiz. }{\bf #1} (19#2) #3}
\def\sjnp#1#2#3{   {\it Sov. J. Nucl. Phys. }{\bf #1} (19#2) #3}
\def\jetp#1#2#3{   {\it Sov. Phys. }{JETP }{\bf #1} (19#2) #3}
\def\jetpl#1#2#3{  {\it JETP Lett. }{\bf #1} (19#2) #3}
\def\ppsjnp#1#2#3{ {\it (Sov. J. Nucl. Phys. }{\bf #1} (19#2) #3}
\def\ppjetp#1#2#3{ {\it (Sov. Phys. JETP }{\bf #1} (19#2) #3}
\def\ppjetpl#1#2#3{{\it (JETP Lett. }{\bf #1} (19#2) #3} 
\def\zetf#1#2#3{   {\it Zh. ETF }{\bf #1}(19#2) #3}
\def\cmp#1#2#3{    {\it Comm. Math. Phys. }{\bf #1} (19#2) #3}
\def\cpc#1#2#3{    {\it Comp. Phys. Commun. }{\bf #1} (19#2) #3}
\def\dis#1#2{      {\it Dissertation, }{\sf #1 } 19#2}
\def\dip#1#2#3{    {\it Diplomarbeit, }{\sf #1 #2} 19#3 }
\def\ib#1#2#3{     {\it ibid. }{\bf #1} (19#2) #3}
\def\jpg#1#2#3{        {\it J. Phys}. {\bf G#1}#2#3}  
\newcommand{\bas}{\bar{\alpha}_S}
%
%
%
%
%
\noindent
\begin{flushright}
\parbox[t]{10em}{
{\bf TAUP-2731-2003}\\
~ \\ 
{\bf \today} }\\
\end{flushright}
\vspace{1cm}
\begin{center}
{{\LARGE  \bf   $\mathbf{\gamma^{*}}$-$\mathbf{\gamma^{*}}$
Scattering: Saturation and}\\
{\LARGE  \bf
 Unitarization in the BFKL Approach.}
\vskip1cm
{\large \bf S. Bondarenko ${}^{a)}$ \footnote{Email:
serg@post.tau.ac.il.}, ~ M. ~Kozlov ${}^{a)}$ \footnote{Email:
kozlov@post.tau.ac.il
.} and E. ~Levin ${}^{a),b)}$\footnote{Email:
leving@post.tau.ac.il, levin@mail.desy.de.}}}
\vskip1cm

{\it ${}^{ a}$)\,\, HEP Department}\\
{\it School of Physics and Astronomy}\\
{\it Raymond and Beverly Sackler Faculty of Exact Science}\\
{\it Tel Aviv University, Tel Aviv, 69978, Israel}\\
\vskip0.3cm
{\it ${}^{b)}$ DESY Theory Group}\\
{\it 22603, Hamburg, Germany}

\end{center}  
\bigskip
\begin{abstract} 	
In this paper $\gamma^* - \gamma^*$ scattering with large, but more or 
less
equal virtualities of two photons  is discussed using  BFKL
dynamics, emphasizing the large impact parameter behavior ($b_t$) of the
dipole-dipole amplitude. It is shown that the non-perturbative  
contribution is essential  to fulfill the unitarity constraints in the
region of $b_t \,>\,1/2\,m_\pi$, where
$m_{\pi}$ is pion mass. The saturation and the unitarization of the
dipole-dipole amplitude is considered in the framework of the
Glauber-Mueller
approach. The main result is that we can satisfy the unitarity constraints
introducing the non-perturbative corrections only in initial conditions
( Born amplitude).

\end{abstract}

\newpage 

\section{Introduction.}
\label{sec:Introduction}
In this paper we continue our investigation of  $\gamma^* - \gamma^*$ 
scattering
at high energies
(see Ref.\cite{KL} for our previous attempts to study this process in the DGLAP dynamics).
We concentrate our efforts here on the case of two photons with large but almost equal
virtualities. It has been argued \cite{BRL,BHS} that this process is the perfect tool to
recover the
BFKL dynamics \cite{BFKL} which is the key problem in our understanding of the low $x$ ( high
energy) asymptotic behavior in QCD.

\begin{figure}[htbp]
\begin{minipage}{10.0cm}
\epsfig{file= 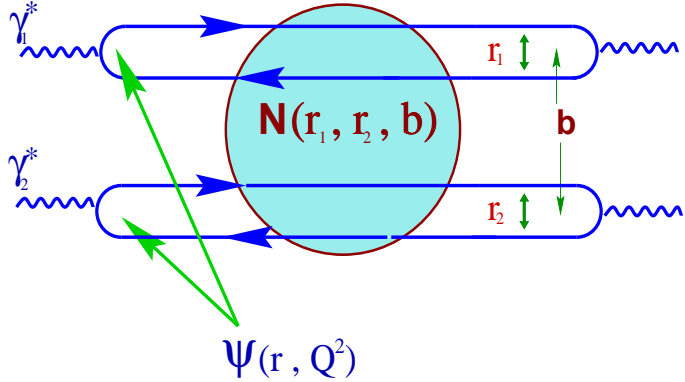,width=90mm, height=50mm}
\end{minipage}
\begin{minipage}{6.0 cm}
\caption{The picture of interaction of two  photons with virtualities
$Q_1$ and $Q_2$ larger than a  ``soft" scale.}
\label{ddint}
\end{minipage}
\end{figure}

It is well known that the correct degrees of freedom at high energy are
not quarks or gluon but  colour dipoles \cite{MU90,LR87,KOP,MU94} 
which
have transverse sizes $r_t$ and the fraction of energy $z$.
Therefore, two photon interactions occur in two successive   steps.
First, each 
virtual photon decays into a colour dipole ( quark - antiquark pair ) with
size $r_t$. At large value of photon virtualities the probability of such a
decay can be calculated in pQCD. The second stage is the interaction of
colour dipoles with each other. The simple formula ( see for example
Ref. \cite{DDR} ) that describes the
process of interaction of two photons with virtualities $Q_1$ and $Q_2$ ($
 \,\approx\,Q_1$ ) 
is  (see \fig{ddint} ) 
\beq \label{PPS}
\si(Q_1, Q_2,W)\,\,=\,\,
\int\,d^2 b_t
\sum^{N_f}_{a,b}\,\
\eeq
$$
\int^1_0 \,d\,z_1 \,\int\,d^2\,r_{1,t} |
\Psi^a_{T,L}(Q_1;z_1,r_{1, t})|^2\,\,\int^1_0 \,d\,z_2 \,\int\,d^2 \,
r_{2,t} |
\Psi^b_{T,L}(Q_2; z_2,r_{2, t})|^2\,\,2\,N(x,r_{1,t},r_{2,t}; b_t)
$$
where the indexes $a$ and $b$ specify the flavors of interacting quarks, 
$T$ and $L$ indicate the polarization of the interacting photons where 
$r_i$ denote the transverse separation between quark and antiquark in the
dipole (
dipole size) and $z_i$ are the energy fractions of the quark in the 
fluctuation of  photon
$i$ into quark-antiquark pair.   $N(x,r_{1,t},r_{2,t}; b_t) $ is the 
imaginary part of the dipole - dipole amplitude 
at   $x$ given by
\beq \label{X}
x \,\,=\,\,\frac{Q^2_1\,+\,Q^2_2 }{W^2\,+\,Q^2_1\,+\,Q^2_2}
\eeq
for massless quarks ($W$ is the energy of colliding photons in c.m.f.). 
$b_t$ is the impact
parameter for dipole-dipole interaction and it is equal the transverse
distance between the dipole centers  of mass. 

The wave functions for virtual photon are  known \cite{WF} and they are
given by (for massless quarks)
\begin{eqnarray}
|\Psi^a_{T}(Q; z,r_{ t})|^2 &=& \sum_a\,\frac{6
\alpha_{em}}{\pi^2}\,Z^2_a\,\,(z^2\,+\,(1 -
z)^2)\,\bar{Q}^2\,K^2_1 (\bar{Q}\,r_t)\,\,;\label{WFT}\\
|\Psi^a_{L}(Q; z,r_{ t})|^2
&=& \sum_a\,\frac{6\alpha_{em}}{\pi^2}\,Z^2_a\,Q^2\,z^2\,(1 - z)^2
K^2_0(\bar{Q}\,r_t)\,\,;\label{WFL}
\end{eqnarray}
with $ \bar{Q}^2_a \,\,=\,\,z(1-z)Q^2\,$   where
$Z_a$  denote the faction of quark charge   of
flavor $a$.

Since the  main contribution in \eq{PPS} is concentrated at $r_{1,t}\,
\approx\,1/Q_1 \,\ll \,1/\mu$ and $ r_{1,t}\,\approx\,1/Q_2\,\ll\,1/\mu$ 
where $\mu$ is the soft mass scale, we can
safely use pQCD for calculation of the dipole-dipole amplitude $N$
in \eq{PPS}.

In this paper  we   study this process in the region of high energy and  
large 
but more-less equal photon virtualities 
($Q^2_1 \,\approx\,\,Q^2_2 \,\,\gg\,\,1/\mu^2$) in the
framework of the BFKL dynamics. In the region of very small
 $x$ (high energies) the saturation
of the gluon density is expected \cite{GLR,MUQI,MV}. We will deal with this phenomenon using
Glauber-Mueller formula \cite{MU90,LR87,KOP} which is the simplest one that reflects all
qualitative features of a more general approach based on non-linear evolution
\cite{GLR,MUQI,MV,KV}.  For $\gamma^* - \gamma^*$ scattering with large but equal photon
virtualities,  the Glauber-Mueller approach is the only one  on the market since
the non-linear equation is justified only for the case when one of the photon has larger
virtuality than the other.

In the next section  we discuss the dipole-dipole
interaction in the BFKL approach  of pQCD.  The solution to the BFKL equation, that
describes
the dipole-dipole interaction in our kinematic region, has been found \cite{LIP} and our
main
 concern in this section is to find the large impact parameter ($b_t$) behavior of the
solution.
As was discussed in Ref. \cite{LRREV,KW,FIIM,KL}, we have to introduce 
non-perturbative  corrections in the region of  $b_t$ larger than $1/2\,m_\pi$ where $m_\pi$
is the pion mass. We argue in this section that it is sufficient to introduce the
non-perturbative behavior into the Born approximation to obtain a reasonable solution at
large $b_t$.

Section 3 is devoted to Glauber - Mueller formula in the case of the BFKL
emission \cite{BFKL}. Here, we use the advantage of  photon - photon
scattering with large photon virtualities, since we can calculate the
gluon
density without uncertainties related to non-perturbative initial
distributions in hadronic target. We consider the low $x$ behavior of the dipole-dipole
cross section and show that the large impact parameter behavior,  introduced in the Born
cross section,  fulfills  the unitarity restrictions ( unitarity bound
\cite{FROI}). Therefore, we confirm that the large $b_t$ behavior can be concentrated in
the initial condition (see Refs. \cite{LRREV,FIIM,KL} without changing the kernel of the
non-linear equation that governs evolution in the saturation region
as it is advocated in Ref.\cite{KW}.

In the last section we summarize our results.

\section{Dipole-dipole interaction in the BFKL approximation.}
\label{sec:DipoleDipoleInteraction}
 
\begin{figure}
\begin{center}
\epsfig{file=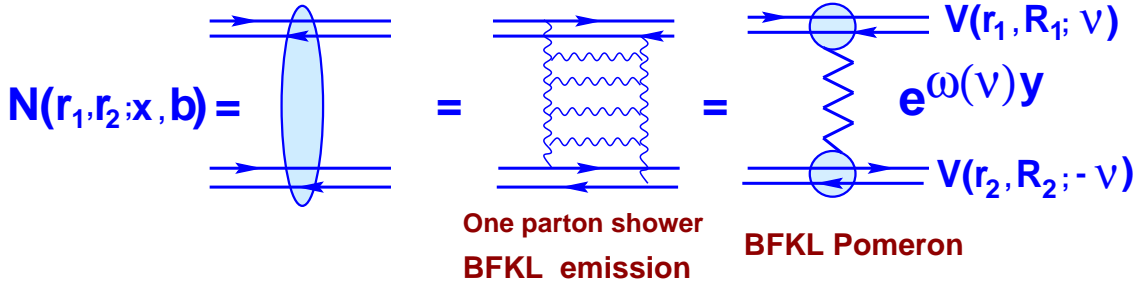,width=150mm}
\end{center}
\caption{One parton shower interaction in the BFKL approach.}
\label{ops}
\end{figure}

In this section we discuss the one parton shower interaction in the BFKL dynamics ( see
\fig{ops}). We start with the Born approximation which is the exchange of
 two gluons (see \fig{badip} ) or  the
diagrams of \fig{ops} without emission of a gluon.

\subsection{ Born Approximation:\,\,\,}

\begin{figure}
\begin{center}
\epsfig{file=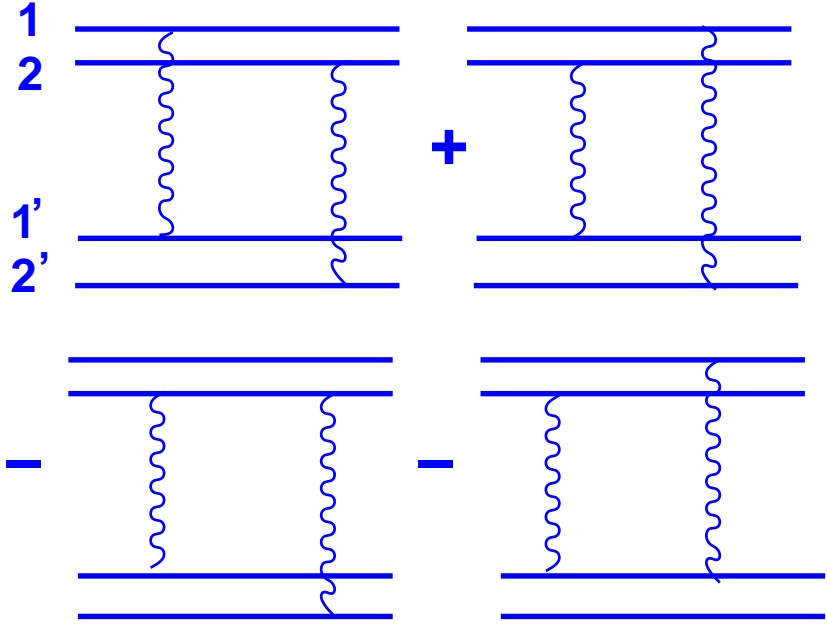,width=110mm}
\end{center}
\caption{Dipole-dipole interaction in the Born approximation.}
\label{badip}
\end{figure}

 These diagrams have been calculated in Ref. \cite{KL} using
the approach of Ref. \cite{BAA} and they lead to the following expression for the 
dipole-dipole
amplitude:

\begin{eqnarray} 
N^{BA}( r_{1, t},r_{2, t}; b_t) \,\,&=& \,\, \,\pi \as^2
\frac{N^2_c - 1}{2\,N^2_c}\,\,\left( \,\ln
\frac{(\vec{b}\,-\,z_1 \vec{r}_1 \,-\, z_2\vec{r}_2  )^2\,\,
(\vec{b}\,-\,\bar{z}_1 \vec{r}_1\, -\, \bar{z}_2 \vec{r}_2 )^2}
{(\vec{b}\,- \,\bar{z}_1 \vec{r}_1 \,- \,z_2  \vec{r}_2 )^2\,\,
(\vec{b}\,- \,z_1 \vec{r}_1\, - \,\bar{z}_2  \vec{r}_2)^2}\,\right)^2
 \label{BAXS} \\
 &=&\,\,\pi \as^2 \frac{N^2_c - 1}{2\,N^2_c}\,\ln^2 
\left(\frac{\rho^2_{1,1'} 
\,\rho^2_{2,2'}}{\rho^2_{1,2'}\,\rho^2_{2,1'}}\right) \label{BAXS1}
\end{eqnarray}  
where $z_i$ is the fraction of the energy of the dipole carried
by quarks; $\bar{z}_i \,=\,z_i\,-\,1$ and $\rho_{i,k} \,=\,\vec{\rho}_i - 
\vec{\rho}_k$. $\vec{\rho}_i $ is the coordinate of quark $i$ (see 
\fig{badip}). All vectors are two dimensional in \eq{BAXS}.

Each diagrams in \fig{badip} is easy to calculate \cite{LI} and the first 
diagram is equal to
\beq \label{BA2}
\pi \as^2 \frac{N^2_c - 1}{2\,N^2_c}\,\ln 
\rho^2_{1,1'}\,\,\ln\rho^2_{2,2'}\,\,.
\eeq
Summing all diagrams we obtain \eq{BAXS}.

We are interested mostly in the limit of large $b_t\,\,\gg\,\,r_{1,t} \,\approx\,\,r_{2,t}$
where the dipole-dipole amplitude can
be reduced to a simple form.
\beq \label{BALB}
N^{BA}( r_{1, t},r_{2, t}; b_t)\,\,\rightarrow\,\,\pi \as^2
 \frac{N^2_c - 1}{ N^2_c}\,\,\frac{r^2_{1, t}\,\,r^2_{2, t}}{b^4_t}\,\,,
\eeq
after integration over azimuthal angles.
 
Therefore, we have a power-like decrease of the dipole-dipole amplitude at large $b_t$ ,
namely $N^{BA}\,\propto\,\frac{r^2_{1,t}\,\,r^2_{2,t}}{b^4_t}$. Such
behavior cannot be correct
since it contradicts  the general postulates  of analyticity and
crossing symmetry of the scattering amplitude \cite{FROI}. Since the spectrum of
hadrons  has no particles with mass zero,  the scattering amplitude should
decrease  as $e^{ - 2m_{\pi}\,b_t}$ \cite{FROI}. In Ref. \cite{KL} we suggested a 
procedure of  how to cure this problem which is based on the results of QCD sum rules
\cite{QCDSR}.Following  this procedure we rewrite the dipole-dipole amplitude as the
integral over the  mass of two gluons in $t$-channel; and we assume, as in QCD sum
rules, that
this integral describes all hadronic states on average. Restricting the integral over
mass
by the minimal mass of  hadronic states ( 2 $m_\pi$ ) we obtain the model which provides
the exponential fall at large $b_t \,\gg\,1/(2\,m_\pi)$ and does not change the power like
behavior for small $b_t\,\ll\,1/(2\,m_\pi)$. 

We choose for the Born amplitude the following formula
\beq \label{BACORR}
N^{BA}(r_{1, t},r_{2, t}; b_t)\,\,=\,\,\pi \as^2 \frac{N^2_c - 1}{
N^2_c}\,\,
\frac{r^2_{1,t}\,r^2_{1,t}\,m^4_\pi}{3}\,K_4(2\,m_\pi b_t) \,\,.
\eeq
One can easily see that \eq{BACORR} reproduces \eq{BALB} and leads to
\beq \label{BACORLB}
N^{BA}( r_{1, t},r_{2, t}; b_t)\,\,\rightarrow\,\,\pi \as^2 \frac{N^2_c
- 1}{ N^2_c}\,\,
\frac{r^2_{1,t}\,r^2_{2,t}\,m^4_\pi}{3}\,\sqrt{\frac{\pi}{2\,m_\pi\,b_t}}\,e^{
-
2\,m_\pi\,b_t} \,\,.
\eeq
at large $b_t \,\,\gg\,\,1/(2\,m_\pi)$.

\subsection{BFKL equation:} 

The emission of a gluon is described by the BFKL equation \cite{BFKL} which was solved in
Ref.\cite{LIP} for fixed $b_t$ (see Ref.\cite{BFLLRW,NP,LI} for many useful discussion of 
the 
different
aspects  of the solution). The solution can be presented in factorized form (see
\fig{ops} ).
\beq \label{BFKL1}
N(x, r_{1, t},r_{2, t}; b_t)\,\,=
\eeq
$$
\int \frac{d \nu}{2\,\pi\,i}\,\phi_{in}(\nu;r_{1,t}; b_t)
\,\,d^2\, R_1 \,\,d^2\,R_2\,\delta(\vec{R}_1 - \vec{R}_2 - \vec{b}_t)\,
e^{\omega(\nu)\,y}
\,V(r_{1,t},R_1;\nu)\,V(r_{2,t},R_2;-\nu)
$$
with
\beq \label{OMEGA}
\omega(\nu)\,\,=\,\, \frac{\as N_c}{\pi}\,\left(\,2 \,\psi(1) \,-\,\psi(\h - i\,\nu)\,-
\,\psi(\h +
i\,\nu)\,\right)\,;
\eeq
 where $\psi(f) \,=\,d \ln \Gamma(f)/d f$,$ \Gamma(f)$ is Euler gamma function and where
\beq \label{V}
V(r_{i,t},R_i;\nu)\,\,=\,\,\left( \,\frac{r^2_{i,t}}{(\vec{R}_i
\,+\,\frac{1}{2}\vec{r}_{i,t})^2\,\,
(\vec{R}_i\,-\,\frac{1}{2}\vec{r}_{i,t})^2}\,\right)^{\frac{1}{2}
\,-\,i\,\nu}
\eeq
using  the following notations: $y\,=\,\ln(x_0/x)$;\,\,$r_{i.t}$  is the size of the colour 
dipole
$``i"$ and $R_i$ is the position of the center of mass of  this dipole.
In \eq{BFKL1}  function $ \phi_{in}(\nu;r_{2,t}; b_t)$ should be found from the initial
condition which
determines the dipole amplitude at fixed $x=x_0$,
namely, $N(x=x_0, r_{1, t},r_{2, t}; b_t) \,\,=\,\,N^{BA}(x=x_0, r_{1, t},r_{2, t}; b_t)$.

It should be stressed that the BFKL equation is a linear equation in which the kernel does 
not 
depend on
$b_t$ (see Ref.\cite{KV}). Therefore, $\phi_{in}(\nu;b_t)$ could be an arbitrary function 
on $b_t$.

In \eq{BFKL1} we can take the integral over $R_2$ which leads to
\beq \label{V2}
V(r_{2,t},R_2;-\nu)\,\,=\,\,\left( \,\frac{r^2_{2,t}}{(\vec{R}_1
\,+\,\vec{b}_t
\,+\,\frac{1}{2}\vec{r}_{2,t})^2\,\,
(\vec{R}_1\,+\,\vec{b}_t\,-\,\frac{1}{2}\vec{r}_{2,t})^2}\,\right)^{\frac{1}{2}
\,+\,i\,\nu}\,\,.
\eeq 

We are interested in the large $b_t$ behavior, namely, $b_t
\,\,\gg\,\,r_{1,t}\,\approx\,\,r_{2,t}$.
It is instructive to consider two cases:
\begin{itemize}
\item\quad {\bf DLA:} 
\begin{boldmath}${\h - i\,\nu\,\,\rightarrow\,\,0}$.
\end{boldmath} This
is
so called double log approximation
of pQCD (DLA) in which we consider $ r_{1,t}\,\ll\,r_{2,t}$ and $\as \ln(1/x)\,\ln
(r^2_{2,t}/r^2_{1,t})\,\,\approx\,1$ while $\as
\ln(1/x)\,\ll\,1$ as well as $\as \ln (r^2_{2,t}/r^2_{1,t}) \,\ll\,1$ and $\as\,\ll\,1$. We have
considered this case
in Ref.\cite{KL} and found  that  the  emission of gluons does not induce  any additional
dependence on
$b_t$ which is concentrated only in the Born amplitude. Indeed, we can see this property 
directly
from the solution of \eq{BFKL1}. 

 Integrating  over $R_1$ we find that the integrand
of 
this integral
falls down rapidly  for  $R_1\,>\,b_t$ due to $R_1$ dependence  of 
the vertex $V(r_{2,t},R_2;\nu)$
(see \eq{V2}) providing a good convergence for the integral.
For $R_1 <b_t$ we can neglect $R_1$ dependence of the vertex $V(r_{2,t},R_2;\nu)$ 
and consider
it as $\left( \frac{r^2_{2,t}}{b^4_t}\right)^{\h + i\,\nu}$ . The integral over $R_1$ of
$V(r_{1,t},R_1,\nu)$ for
$R_1 <b_t$ gives $\,(r_{1,t})^{\h -i\,\nu}\,(b^2_t)^{ 2\,i\,\nu}\,$.

 Therefore, $ \int d^2 R_1
V(r_{1,t},R_1,\nu)\,\,V(r_{2,t},b_t,\nu)\,\,\rightarrow\,
\,(r^2_{2,t}/b^2_t) \,(r^2_{1,t}/r^2_2)^{\h -
i\,\nu}$. Finally, 
taking $\phi_{in}(\nu;r_{1,t}; b_t)\,=
\,\pi\, \as \,\frac{N^2_c - 1}{3\,\,N^2_c}\,\,(m_\pi\,)^2\,(r_{1,t}\,b_t)^2\,\,K_4(
2\,m_\pi\,b_t)\,\frac{1}{\ \h \,-\,i\,\nu}\,\,$,
the dipole amplitude has a form
\beq \label{DLA1}
N^{DLA}(x, r_{1, t},r_{2, t}; b_t) \,=\,N^{BA}(x, r_{1, t},r_{2, t}; b_t)\,\int \,\frac{d \nu}{2
\,\pi\, i}\,\,e^{\omega(\nu) \,y \,\,+\,\, (\h \,-\,i\,\nu)\,\ln(r^2_{1,t}/r^2_{2,t})}
\eeq
Considering  $r_{2,t}\,\ll\,r_{1,t}$ and taking into account that $\omega(\nu)\,\,\rightarrow
\frac{\as N_c}{\pi}\,\frac{1}{\h \,-\,i\,\nu}$ at $\h \,-\,i\,\nu \,\rightarrow \,0$ one 
can take 
the integral in \eq{DLA1} explicitly. The answer is well known ( see Ref.\cite{KL} for example),
namely,
at low $x$
\beq \label{DLA2}
N^{DLA}(x, r_{1, t},r_{2, t}; b_t)\,\,=\,\,N^{BA}(x, r_{1, t},r_{2, t}; b_t)\,\,I_0 \left(2
\sqrt{\frac{\as\, N_c}{\pi}\,y\,\ln(r^2_{2,t}/r^2_{1,t})}\,\right)
\eeq
for  fixed coupling constant\footnote{In this paper we consider only the case of fixed QCD 
coupling since the BFKL equation is not proven for running $\as$.}.

\item \quad {\bf Diffusion approximation:}
\begin{boldmath} ${\nu\,\,\ll\,\,1}$. \end{boldmath} For such small values
of $\nu$ the  integral over $R_1$ is convergent for  $R_1 \,> r_{1,t}$ (see
Ref.\cite{BFLLRW} ) and,
therefore, we  neglect
the $R_1$ dependence in $V(r_{2,t},b_t,\nu)$. Introducing a new variable $\vec{\xi}  =
\vec{R}_1/r_{1,t}$ we  see that 
\beq \label{DF1}
\int d^2 \,R_1 V(r_{1,t},R_1;\nu) \,=\,(r^2_{1,t})^{\h + i\,\nu}\,
\int d^2 \xi \left( \vec{\xi} +
\h\,\vec{n})^2\,( \vec{\xi} - \h\,\vec{n})^2\,\right)^{\h + i\,\nu}
\eeq
where $\vec{n}$ is a unit vector in the direction of $\vec{r}_{1,t}$. The integral
is a function of $\nu$ only and can  be absorbed in $\phi_{in}(\nu; b_t)$ in \eq{BFKL1}.
For $V(r_{2,t},b_t,-\nu)$ at $b_t\,\gg\,r_{2,t}$ we have  
\beq \label{VDF}
V(r_{2,t},b_t,-\nu)\,\,=\,\,\left(\frac{r^2_{2,t}}{b^4_t}\right)^{\h +
i\,\nu}
\eeq
Therefore, the dipole amplitude is
\beq \label{NDF}
N^{DF}(x, r_{1, t},r_{2, t}; b_t)\,\,=\,\,\int \,\frac{d \nu}{2\,\pi\,i}\,
\phi_{in}(\nu; b_t)\,\,
e^{\omega(\nu)\,y} \,\,\left(\,\frac{r^2_{1,t}\,r^2_{2,t}}{b^4_t}\,\right)^{\h + i\,\nu}\,\,.
\eeq
We  choose $\phi_{in}(\nu; b_t)$ to be of  in the form
\beq \label{INCON}
\phi_{in}(\nu; b_t)\,\,=\,\,\pi\, \as \,\frac{N^2_c - 1}{3\,\,N^2_c}\,\,(m_\pi\,b_t)^4\,\,K_4(
2\,m_\pi\,b_t)\,\frac{1}{\ \h \,-\,i\,\nu}\,\,.
\eeq
At small values of $\nu$ we can expand $\omega(\nu)$  
\beq \label{KERNDF}
\omega(\nu)\,\,=\,\,\omega_L\,\,-\,\,D\,\nu^2
\eeq
with
\beq \label{OMSN}
\omega_L\,\,=\,\,\frac{\as\,N_c}{\pi}\,4\,\ln2\,\,;\,\,\,\,\,\,\,\,\,\,\,
D\,\,=\,\,\frac{\as\,N_c}{\pi}\,\,14\,\zeta(3)\,\,;
\eeq
Finally, we can evaluate the  integral over $\nu$ in \eq{NDF} using the method of  steepest 
decent  and obtain
the following expression for dipole amplitude:
\beq \label{ANDF}
N^{DF}(x, r_{1, t},r_{2, t}; b_t)\,\,= 
\eeq
$$ \pi\, \as \,\frac{2(N^2_c -
1)}{3\,\,N^2_c}\,\,(\,r_{1,t}\,r_{2,t}\,m^4_\pi
\,b^2_t\,)\,K_4(2\,m_\pi\,b_t)\,\,\sqrt{\frac{\pi}{D\,y}}\,\,e^{
\omega_L\,y \,\,-\,\,\frac{\ln^2\frac{r^2_{1,t}\,r^2_{2,t}}{b^4_t}}{4\,D\,y}} 
$$

At $ 1/(2\,m_\pi) >\,b_t\,>\,r_{1,t} \,\approx\,r_{2,t}$
\beq \label{ANDFB1}
N^{DF}(x, r_{1, t},r_{2, t}; b_t)\,\,\rightarrow
\eeq
$$
  \pi\, \as \,\frac{2(N^2_c -
1)}{3\,N^2_c}\,\,\frac{r_{1,t}\,r_{2,t}}{b^2_t}\,\sqrt{\frac{\pi}{D\,y}}\,\,e^{
\omega_L\,y \,\,-\,\,\frac{\ln^2\frac{r^2_{1,t}\,r^2_{2,t}}{b^4_t}}{4\,D\,y}}
$$

while at $b_t\,>\,1/(2\,m_\pi)$
\beq \label{ANDFB2} 
N^{DF}(x, r_{1, t},r_{2, t}; b_t)\,\,\rightarrow 
\eeq
$$
   \pi\, \as \,\frac{2(N^2_c -
1)}{3\,N^2_c}\,\,(\,r_{1,t}\,r_{2,t}\,m^4_\pi
\,b^2_t\,)\,\sqrt{\frac{\pi}{2 \,m_\pi\,b_t}} \sqrt{\frac{\pi}{D\,y}}\,\,e^{
\omega_L\,y
\,\,-\,\,\frac{\ln^2\frac{r^2_{1,t}\,r^2_{2,t}}{b^4_t}}{4\,D\,y}\,\,-\,\,2\,m_\pi\,b_t}\,\,\,
$$

\end{itemize}
 
\section{Saturation and unitarization in the Glauber - Mueller approach.}

\subsection{ Glauber - Mueller formula.}
\begin{figure}[htbp]
\begin{minipage}{10.0cm}
\epsfig{file= 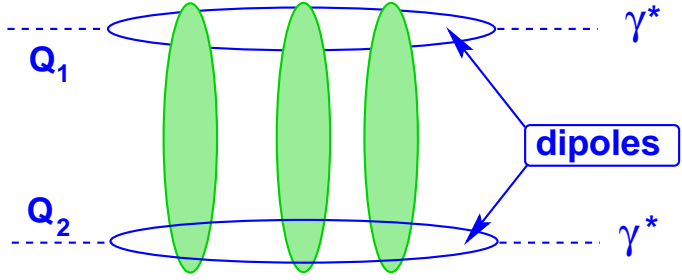,width=90mm}
\end{minipage}
\begin{minipage}{6.0 cm}
\caption{The Glauber - Mueller  approach for the dipole-dipole scattering
amplitude.}
\label{gm}
\end{minipage}
\end{figure}

The Glauber - Mueller approach \cite{MU90,LR87,KOP} takes into account the interaction 
of many
parton showers with the target as  is shown in \fig{gm}. In our case of more or less 
equal but
large virtualities of both photons  this approach gives a unique opportunity to study the high
energy asymptotic behavior of the dipole amplitude since other methods based on non-linear
evolution equation \cite{GLR,MUQI,MV,KV,SAT,ELTHEORY} do not work in the case of two dipoles
with more-less equal sizes.

The main idea of this approach is that the  colour dipoles  are  the correct
degrees of freedom for high energy scattering (this idea was
formulated by A.H. Mueller in Ref. \cite{MU94}). Indeed, the
change of the value of the dipole size $r_t$ ($\Delta r_t$) during the
passage of the colour dipole through the target is proportional to
the number of rescatterings (or the size of the target $R$) multiplied by
the angle $k_t/E$ where $E$ is the energy of the dipole and $k_t$ is the
transverse momentum of the $t$-channel gluon which is emitted by the fast
dipole.

\beq \label{DOF1}
\Delta \,r_t \,\,\propto\,\,R\,\frac{k_t}{E}\,\,.
\eeq
Since $k_t$ and $r_t$ are conjugate variables and due to the uncertainty
principle 
$$
k_t\,\,\propto\,\,\frac{1}{r_t}\,\,.
$$
 
Therefore,
\beq \label{DOF2}
\Delta\,r_t\,\,\propto\,\,R\,\frac{k_t}{E}\,\,\ll\,\,r_t\,\,\,
\mbox{if}\,\,\,
\,R\,\ll\,\,r^2_t\,\,E\,\,\,\mbox{or}\,\,\,x\,\,\ll\,\,\frac{1}{2
m\,R}\,
\,.
 \eeq

Since the colour dipoles are correct degrees of freedom , they diagonalize the interaction
matrix at high energy as well as the unitarity constraints, which have the form
\beq \label{UN}
2\,\,N (x,r_{1,t},r_{2,t};b_t)\,\,=\,\,|a_{el}
(x,r_{1,t},r_{2,t};b_t)|^2\,\,+\,\,G_{in}(x,r_{1,t},r_{2,t};b_t)\,\,,
\eeq
where $a_{el}$ is the elastic amplitude of the dipole-dipole interaction and $N$ is the
imaginary part of $a_{el}$ ($N=Im \,a_{el}$).

Assuming that the amplitude is pure imaginary at high energy, one can find a simple solution
to \eq{UN}, namely
\begin{eqnarray}
a_{el}(x,r_{1,t},r_{2,t};b_t)\,\,&=&\,\,i \,\left(
1\,\,-\,\,e^{- \frac{\Omega(x,r_{1,t},r_{2,t};b_t)}{2}}\,\right)\,\,;
\label{UNTEL}\\
G_{in}(x,r_{1,t},r_{2,t};b_t)\,\,&=&\,\,\left(
1\,\,-\,\,e^{- \Omega(x,r_{1,t},r_{2,t};b_t)}\,\right)\,\,;
\label{UNTIN}
\end{eqnarray}
where $\Omega$ is the arbitrary real function.

In Glauber - Mueller approach the opacity $\Omega$ is chosen as $\Omega =
2\,N^{OPS}(x,r_{1,t},r_{2,t};b_t)$ where $N^{OPS}$ is the dipole-dipole amplitude for one
parton shower interaction that has been found in the previous section (see \eq{BFKL1}). 

\subsection{ Saturation.} 

One can see that if we substitute the explicit solution
to the BFKL equation
of \eq{ANDF} at any fixed $b_t$ the opacity $\Omega = 2 N^{DF}$ increases at $x
\,\rightarrow\,0$. Therefore, the dipole-dipole amplitude given by Glauber-Mueller
formula of \eq{UNTEL} tends to unity in the region of low $x$. This statement is called 
saturation \cite{GLR,MUQI,MV} since the physical interpretation of $N$ is the density of
colour dipoles at least when $N$ is not very large. In this discussion the saturation 
appears to be  
the consequence of  unitarity for fixed $b_t$. However, we have learned several examples
where  the dipole density could reach a maximum value without having any effect on the 
elastic
dipole-dipole amplitude at fixed $b_t$ (see Ref. \cite{MV} and paper of Kovchegov and
Mueller in Ref.\cite{SAT}). However, for $\gamma^* - \gamma^*$ scattering of two small dipoles
the
initial condition is given by Born amplitude of \eq{BAXS} which is small.   Therefore, we have
no reason to expect that the dipole density will be high due to the final state interaction.

\subsection{ Unitarization.}

 To obtain the unitarity bound for the dipole-dipole cross 
section
we have to integrate over $b_t$, namely
\beq \label{DDXS}
\sigma(dipole - dipole)\,=\,2\int d^2\,b_t\,N^{GM}(x,r_{1,t},r_{2,t};b_t) =
2\int d^2\,b_t\,\left(1\,-\,e^{- N^{DF}(x,r_{1,t},r_{2,t};b_t)} \right)
\eeq

Following Froissart \cite{FROI}, we  divide the region of integration over $b_t$ in
\eq{DDXS} in two parts
\beq \label{DDUN1}
\sigma( dipole - dipole)\,\,=\,\,2 \pi \int^{b^2_0(x)}_0
\,d\,b^2_t\,\,N^{GM}(...;b_t)\,\,+\,\,\int^{\infty}_{b^2_0(x)} \,d b^2_t\, \,N^{GM}(...;b_t)
\eeq
where $b_0(x)$ is defined from the equation
\beq \label{b0}
N^{DF}(x,r_{1,t},r_{2,t};b_0(x))=1\,\,\,
\eeq

It is easy to see that for $b_t \,<\,b_0(x)$ $N^{GM} \,\leq\,1$ since $N^{DF} \,>\,1$, while
for $b_t\,\,\geq\,\,1$ and for $N^{DF} \,<\,1$\,\, $N^{GM} \,\leq\,N^{DF}$ 
Therefore, we have the following unitarity bound

\beq \label{DDUN2}
\sigma( dipole - dipole)\,\,\leq\,\,2\,\pi\,\left(\,b^2_0(x)\,\,+\,\,\int^{\infty}_{b_0(x)}\,d
b^2_t
\,N^{GM}(...;b_t)\,\right)
\eeq

Let us consider two possibilities. The first one that $b_0(x)\,\,\ll\,1/(2\,m_\pi) $. In this
case the solution to \eq{b0} follows directly from \eq{ANDFB1} for the amplitude $N^{DF}$ and 
for $y \,\gg\,1$ 
\beq \label{b0s}
\ln (\frac{r^2_{1,t}\,r^2_{2,t}}{b^4_0(x)})\,\,=\,\,-2\,\sqrt{\,D\,\omega_L}\,y
\eeq
or
\beq \label{b0s1}
b^2_0(x)\,\,\propto\,\,r_{1,t}\,r_{2,t}\,\,e^{\sqrt{\,D\,\omega_L}\,y}
\eeq 
Substituting \eq{b0s1} into \eq{DDUN2} we can obtain 
\beq \label{UN3}
\sigma( dipole - dipole)\,\,\leq\,\,2\,\pi \,b^2_0(x)\,\{\,1
\,+\,2\,\}\,\,\propto\,\,
e^{\sqrt{\frac{\,D\,\omega_L}{2}}\,y}
\eeq
where the second term is calculated  by integrating first over $b_t$  \eq{NDF}
and
after that using
saddle point approach. Since $\nu_{SADDLE}$ turns out to be small at low
$x$ and  we neglect it.

 Therefore, in this kinematic region we face a  
power -like increase of
the dipole-dipole cross section as was pointed out in Refs.
 \cite{KW}. 

However, this power-like increase will  stop for $b_0(x) \,>\,1/2\,m_\pi$. Indeed, for
such large values of $b_t$ we should use \eq{ANDFB2} for the dipole amplitude $N^{DF}$.
For such large values of $b_0(x)$ \eq{b0} has a solution which at low $x$  is
\beq \label{SOL1}
b_0(x) \,\,=\,\,\frac{\omega_L }{2 m_\pi}\,y \,\,+\,\,O(ln y)
\eeq 
 which leads to 
\beq \label{UN4}
\sigma( dipole - dipole)\,\,\leq\,\,2\,\pi \,b^2_0(x)\,\,=\,\,\frac{2 \pi\,\omega_L^2}{4
\,m^2_\pi}\,\ln^2(x_0/x)
\eeq 
which comes from the first term in \eq{DDUN2}. It is easy to understand that the second term
in this equation gives a term which does not increase with $y$. \eq{UN4} is the unitarity
bound which has the same energy dependence as for hadron-hadron collisions \cite{FROI} but in
our approach we are able to calculate the coefficient in front of $y^2$. The bound of \eq{UN4}
is the same as was derived in \cite{LRREV,FIIM}.

It should be stressed that the diffusion approximation that we used was derived only at small
values of saddle point in $\nu$ integration in \eq{NDF} which is equal to
\beq \label{DADDLE}
| \nu_{SADDLE}| \,\,=\,\,
\frac{\ln\frac{r^2_{1,t}\,r^2_{2,t}}{b^4_0(x)}}{2\,D\,y}\,\,\ll\,\,1
\eeq
at $b_t=b_0(x)$ from \eq{SOL1}.

\subsection{ Saturation scale.}  
\eq{b0} does not have a solution at any values of $r_{1,t}$ and
$r_{2,t}$ (formally, we obtain a negative values of $b_0(x)$). The same
equation at $b_t = 0$, namely
\beq \label{SATSC}
N(x,r_{sat},r_{2,t};b_t = 0)\,\, = \,\,1\,\,,
\eeq
determines the saturation scale. At $r_{1,t}\, \geq \,r_{sat}$ the
opacity
$\Omega$ in Glauber-Mueller formula is larger than unity ($\Omega
\,\geq\,1$), \eq{b0} has a solution and we are in the saturation region
with \eq{UN4} for the unitarity bound. If $r_{1,t}\, \leq\,r_{sat}$, 
opacity $\Omega \,<\,1$ at any value of $b_t$. This is a domain of
perturbative QCD  in  virtual photon scattering.

$N(x,r_{1,t},r_{2,t};b_t = 0)$ we can find from \eq{BFKL1} integrating over
$R_1$, namely
\beq \label{BFKLb0}
N(x, r_{1, t},r_{2, t}; b_t=0)\,\,=
 \int \frac{d \nu}{2\,\pi\,i}\,\phi_{in}(\nu; b_t = 0)
\,\,d^2\, R_1 \,\,e^{\omega(\nu) y}
\,V(r_{1,t},R_1;\nu)\,V(r_{2,t},R_1;-\nu)\,\,.
\eeq
Since $r_{sat}$ from \eq{SATSC} is much smaller than $r_{2,t}$ we need to
find \eq{BFKLb0} only for $r_{1,t}\,\ll\,r_{2,t}$. This observation
simplifies the calculations. Indeed, the main contribution in the integral
over $R_1$ stems from $R_1 \ll r_{2,t}$. Therefore, we can neglect the
$R_1$ - dependence of vertex $V(r_{2,t},R_1;-\nu)$ which has the form
\beq \label{V2S}
V(r_{2,t},R_1;-\nu) \,\,=\,\,\left(\frac{16}{r^2_{2,t}}\right)^{\h +
i\,\nu}\,\,.
\eeq
To perform the integration over $R_1$ we use the following formula ( see 
equation 3.198 of Ref.\cite{RY} )
\beq \label{FP}
B(\h - i \nu, \h - i \nu)\, \left( \,\frac{1}{(\vec{R}_i
\,+\,\frac{1}{2}\vec{r}_{i,t})^2\,\,
(\vec{R}_i\,-\,\frac{1}{2}\vec{r}_{i,t})^2}\,\right)^{\frac{1}{2}
\,-\,i\,\nu}\,\,=
\eeq
$$
\,\,\int^1_0 \,d\,t\,\, (\,t (1 - t)\,)^{- \h - i\,\nu}
\left( \,R^2_1 \,+\,(1 - 2t) \vec{R}_1\,\cdot\,\vec{r}_{1,t}
\,+\,\frac{r^2_{1,t}}{4}\,\right)^{ - 1 + 2i\,\nu}\,\,,
$$
 where $B(\mu,\nu)= \Gamma(\mu) \Gamma(\nu) /\Gamma(\mu + \nu)$ is the Euler beta -
function. 
Integrating \eq{BFKLb0} over $R_1$ using \eq{FP} we obtain that
\beq \label{BFKLb01}
N(x, r_{1, t},r_{2, t}; b_t=0)\,\,=
 \int \frac{d \nu}{2\,\pi\,i}\,\phi_{in}(\nu; b_t)
 \,\,e^{\omega(\nu) y} \,\frac{B(\h + i\nu,\h + i\nu)}{B(\h - i\nu,\h -
i\nu)}\,\,\frac{1}{ 2\,i\,\nu} \left(\frac{16\,
r^2_{1,t}}{r^2_{2,t}}\right)^{\h + \,i\,nu}
\eeq

The Born approximation at $b_t=0$ and at $r_{2,t}\,\,\geq\,r_{1,t}$ can be
reduced to \cite{KL}
\beq \label{BAb0}
N^{BA}(r_{1, t},r_{2, t}; b_t=0)\,\,\rightarrow \,\,\pi \as^2\,\frac{N^2_c
-1}{N^2_c}\,\,\frac{r^2_{1,t}}{z^2_2\,\tilde{z}^2_2\,\,r^2_{2,t}}\,\,.
\eeq

It is easy to choose $\phi_{in}(\nu; b_t=0)$ in such a way that the final 
answer for $N(x, r_{1, t},r_{2, t}; b_t=0$  is:
\beq \label{FINb0}
N(x, r_{1, t},r_{2, t}; b_t=0)\,\,= \pi
\as^2\,\frac{N^2_c-1}{N^2_c\,z^2_2\,\tilde{z}^2_2} \int
\frac{d \nu}{2\,\pi\,i}\,\frac{1}{\h - i\nu} \,,e^{\omega(\nu)
y\,\,+\,\,( \h + i\nu) \,\ln(r^2_{1,t}/r^2_{2,t})}\,\,.
\eeq

We can find the solution to \eq{SATSC} in the saddle point approximation
for the integral over $\nu$ in \eq{FINb0} \cite{GLR,MUT}. Introducing a new
variable $\gamma
= \h + i \nu$ we have the following equation for the saddle point value of
$\gamma = \gamma_S$:
\beq \label{SADGA}
\frac{d \omega(\gamma)}{d \gamma}|_{\gamma = \gamma_S} \,y
+\,ln(r^2_{1,t}/r^2_{2,t})\,\,=0
\eeq
Substituting \eq{SADGA} in \eq{FINb0} we obtain
\beq \label{SATSC1}
N(x, r_{1, t},r_{2, t}; b_t=0)\,\,\propto\,\,e^{ \omega(\gamma_S)\,y\,-
\gamma_S \ln(r^2_{2,t}/r^1_{2,t})}\,\,=\,\,e^{ y \,\{\omega(\gamma_S)\,\,-
\,\, \gamma_S\,\frac{d \omega(\gamma)}{d \gamma}|_{\gamma = \gamma_S}\}}
\eeq
Using \eq{SATSC1} we can solve \eq{SATSC} in semiclassical approximation ( see Ref. 
\cite{GLR,SAT}  in which we 
cannot calculate the numerical factor in front of \eq{SATSC1} .
Indeed, $N(x, r_{1, t},r_{2, t}; b_t=0)$ is constant on the line
\beq \label{CRL}
\frac{d \omega(\gamma)}{d \gamma}|_{\gamma = \gamma_0} \,y
+\,ln(r^2_{sat}(x)/r^2_{2,t})\,\,=0
\eeq
with $\gamma_0$ is the solution to the equation \cite{GLR,MUT} \footnote{$\gamma_0$ was
called $k_0$ in Ref.\cite{GLR} and $\lambda_0$ in Ref.\cite{MUT}. The
numerical solution of
\eq{CRG} leads to $\gamma_0 \,=\,0.63$.}
 \beq \label{CRG}
\frac{\omega(\gamma_0)}{\gamma_0}\,\,=\,\,\frac{d \omega(\gamma)}{d
\gamma}|_{\gamma = \gamma_0}
\eeq

\eq{CRL} leads to a power-like increase of the saturation momentum 
($ Q_{sat}(x) = 2/r_{sat}$ )
at high energies ( low $x$). Namely,
\beq \label{SATSC2}
Q^2_{sat}(x)\,\,\propto\,\,\frac{1}{r^2_{2,t}}\,
\left(\frac{1}{x}\right)^{\frac{\omega(\gamma_0)}{\gamma_0}}\,\,\approx
\,Q^2_2 \,\,\left(\frac{1}{x}\right)^{\frac{\omega(\gamma_0)}{\gamma_0}}
\eeq 
Actually, the pre-exponential factors in the steepest decent method of taking
integral over $\gamma$  could change the $x$-dependence of
the saturation scale adding some log(1/x) dependence in \eq{SATSC2} ( see
Ref. \cite{MUT} for an  analysis of such corrections).

\subsection{ Unitarity bounds for $\mathbf{\gamma^*}\,-\,\mathbf{\gamma^*}$
scattering.}

To obtain the unitarity bounds for $\gamma^* - \gamma^*$ scattering we
need to substitute the unitarity bound for dipole-dipole cross section
(see \eq{UN4}) into \eq{PPS} and to perform integrations over $r_{i,t}$ and
$z_i$. $ \int \,d^2\,r_t |\Psi_L(Q,z,r_t)|^2 $ is convergent while $ \int
\,d^2\,r_t |\Psi_T(Q,z,r_t)|^2 $ has a logarithmic divergence that we need
to deal with. \eq{UN4} holds only for $r_{1,t}\,>\,r_{sat}$ since if
$r_{1,t}\,<\,r_{sat}$ dipole-dipole cross section is small and
proportional to $\int\,d^2\,b_t N^{OPS}$. As has been mentioned we
consider $r_{1,t}\,\leq\,r_{2,t}$.
On the other hand $K_1(z)
\,\approx\,\,1/z$ at $ z \,<\,1$.
Finally, one can see 
\beq \label{INT}
 \int^{1/\bar{Q_1}}_{r_{sat}}
\,d^2\,r_t\, \,\int^1_0\,d\,z_1 |\Psi_T(Q,z,r_t)|^2\,\,=\,\,
C_Q\, \frac{4}{3}\,\ln (Q^2_{sat}(x)\,r^2_{2,t})\,\,,
\eeq
where $C_Q \,=\,\sum_a\,\frac{6\,\alpha_{em}}{\pi}\,Z^2_a$.
We recall that $Q^2_{sat}(x)\,r^2_{2,t}$ does not depend on $r_{2,t}$.

The integration over $r_{2,t}$ is concentrated at  the limits $ r_{1,t}
\,\leq\,r_{2,t}\,\leq\,1/\bar{Q_2}$ and leads to
 $$\int \,d^2
r_{2,t}\,d\,z_2\,|\Psi_T|^2 \,=\,C_Q\, \frac{4}{3} \,\ln
\frac{1}{r_{1,t}\,Q^2_2} \,\,.$$

Finally, for $\gamma^* - \gamma^*$ cross sections we have
\begin{eqnarray}
\sigma_{T,T}(\gamma^* - \gamma^*) &\leq & C^2_Q
\frac{16}{9}\,\left(
\ln\frac{Q_{sat}(x)}{Q^2_1}\,\ln\frac{Q_{sat}(x)}{Q^2_2}\right)\,\,\,\left(
\frac{\pi\,\omega_L}{
\,m^2_\pi}\,\ln^2(x_0/x)\right)\,\,;\label{XSTT}\\
\sigma_{T,L}(\gamma^* - \gamma^*) &\leq & C^2_Q
\frac{16}{9}\,\left(\ln\frac{Q_{sat}(x)}{Q^2_1}\right)\,\,\,\left(
\,\frac{\pi\,\omega_L}{2  
\,m^2_\pi}\,\ln^2(x_0/x)\right)\,\,;\label{XSTL}\\
\sigma_{L,T}(\gamma^* - \gamma^*) &\leq & C^2_Q
\frac{16}{9}\,\left(\ln\frac{Q_{sat}(x)}{Q^2_2}\,\right)\,\,\,\left(
\frac{\pi\,\omega_L}{2
\,m^2_\pi}\,\ln^2(x_0/x)\right)\,\,;\label{XSLT}\\
\sigma_{L,L}(\gamma^* - \gamma^*) &\leq & C^2_Q
\frac{16}{9}\,\,\left(\frac{\pi\,\omega_L}{2
\,m^2_\pi}\,\ln^2(x_0/x)\,\right)\,\,;\label{XSLL}  
\end{eqnarray}

Since the saturation scale increases as a power of $(1/x)$ one can see
that the energy behavior of the unitarity constraints is 
\begin{eqnarray}
\sigma_{T,T}(\gamma^* -
\gamma^*)\,&\leq\,& \left(C^2_Q\,\frac{16}{3}\right)\,\,
\left(\frac{\pi
\,\omega_L\,(\frac{\omega(\gamma_0)}{\gamma_0})^2}{m^2_{\pi}}\,\right) \,\ln^4(x_0/x)\,\,;
\label{EBTT}\\
\sigma_{T,L}(\gamma^* -
\gamma^*)\,&\leq &  \left(C^2_Q\,\frac{16}{3}\right)\,\,\left(\frac{\pi 
\,\omega_L\,(\frac{\omega(\gamma_0)}{\gamma_0})}{2\,m^2_{\pi}}\,\right) \,\ln^3(x_0/x)\,\,;
\label{EBTL}\\
\sigma_{L,L}(\gamma^* 
-\gamma^*)\,&\leq\,&\left(C^2_Q\,\frac{16}{3}\right)\,\,
\left(\frac{\pi 
\,\omega_L\,}{2\,m^2_{\pi}}\,\right)\,\ln^2(x_0/x)\,\,.
\label{EBLL} 
\end{eqnarray}

Note  only $\sigma_{L,L}(\gamma^* -\gamma^*)$ has the same
energy dependence  as hadron-hadron collisions \cite{FROI} but even this cross section
has a different coefficient in front. $C_Q$ as well the as numerical factor
$2/3$ come from the photon wave function while  $\omega_L$ reflects the BFKL
dynamics making \eq{EBLL} and \eq{XSLL}  quite different from the
unitarity bound for hadronic reactions.

\section{Conclusions.}

In this paper we use $\gamma^* -\gamma^*$ scattering as the laboratory
for studying the large impact parameter behavior of the amplitude in the
saturation
region. At first sight, this processes  occurs at short distances for both
photons with large
virtualities   and could be calculated in
perturbative QCD. We demonstrated that the non-perturbative QCD corrections
have to be introduced for large $b_t$ even for  this process. The main
result of this paper is the statement that it is enough to include the non-perturbative QCD 
corrections   in
the Born approximation and neglect them in the kernel of the BFKL
equation. This result confirms the mechanism suggested in Refs.
\cite{LRREV,FIIM} but it contradicts  the arguments of Ref.
\cite{KW}. 

\begin{figure}[htbp]
\begin{minipage}{10.0cm}
\epsfig{file= 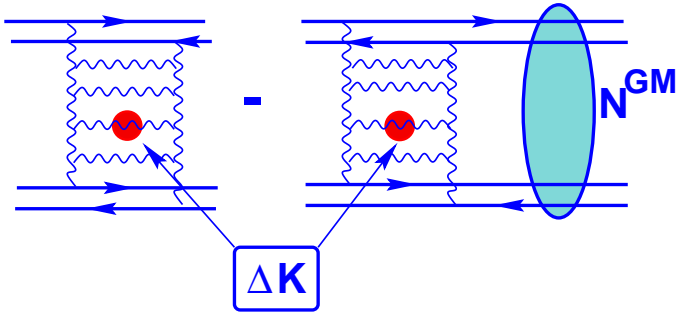,width=90mm}
\end{minipage}
\begin{minipage}{6.0 cm}
\caption{The diagrams for the first order correction with respect to
$\Delta
K$,which includes the non-perturbative QCD contribution at large values
of the impact parameter.}
\label{deltak}
\end{minipage}
\end{figure}

This result does not mean  that the BFKL kernel
correctly describes  the large $b_t$ behavior. The
uncertainties in the large $b_t$ tail of the kernel will not affect the high
energy asymptotic behavior of the dipole amplitude. Let us assume that 
kernel of the BFKL equation can be written  as $K + \Delta K$ where $K$ is
normal BFKL kernel in pQCD and $\Delta K$  includes the non-perturbative
contribution.  We know that $\Delta K \,\propto\,e^{- 2 \,m_\pi\,b_t}$
from  general properties of the strong interaction \cite{FROI}. 
Let us treat $\Delta K$ as a small correction and calculate the first
digram of the order of $\Delta K$ (see \fig{deltak}).

The sum  of all diagrams in \fig{deltak} leads to a contribution 
\beq \label{DLTK}
\Delta K \,\left( 1\,\,-\,\,N^{GM}(x,r_{1,t},r_{2,t},b_t)
\,\right)\,\,=\,\,\Delta K \,e^{ - N^{OPS}(x,r_{1,t},r_{2,t},b_t)}
\eeq
Since for $b_t \,<\,b_0(x)$ $N^{GM}$ is very close to unity, the above
corrections are suppressed. Only for $b_t\,\geq\,b_0(x)$ we can expect a
considerable contribution. However, this contribution is proportional to
$e^{ - 2 \,m_\pi \,b_0(x)} \,=\,e^{ - \omega_L \ln (x_0/x)}$ . Therefore
they turn out to be very small. 

This simple discussion shows  why the strategy   to include the
non-perturbative corrections in the Born amplitude, works

Actually, the main result of this paper, namely \eq{SOL1}, is based on
  a simple
physics  (see Ref. \cite{LRREV}). We have demonstrated here that
the multi rescattering processes embraced by the  Glauber-Mueller formula lead to
a different resulting parton cascade than is given by the BFKL approach.
The principle difference is the fact that the multi parton shower
interaction creates a new scale or mean parton  transverse momentum (
saturation scale) given by \eq{SATSC2}. 

\begin{figure}[htbp]
\begin{center}
\epsfig{file= 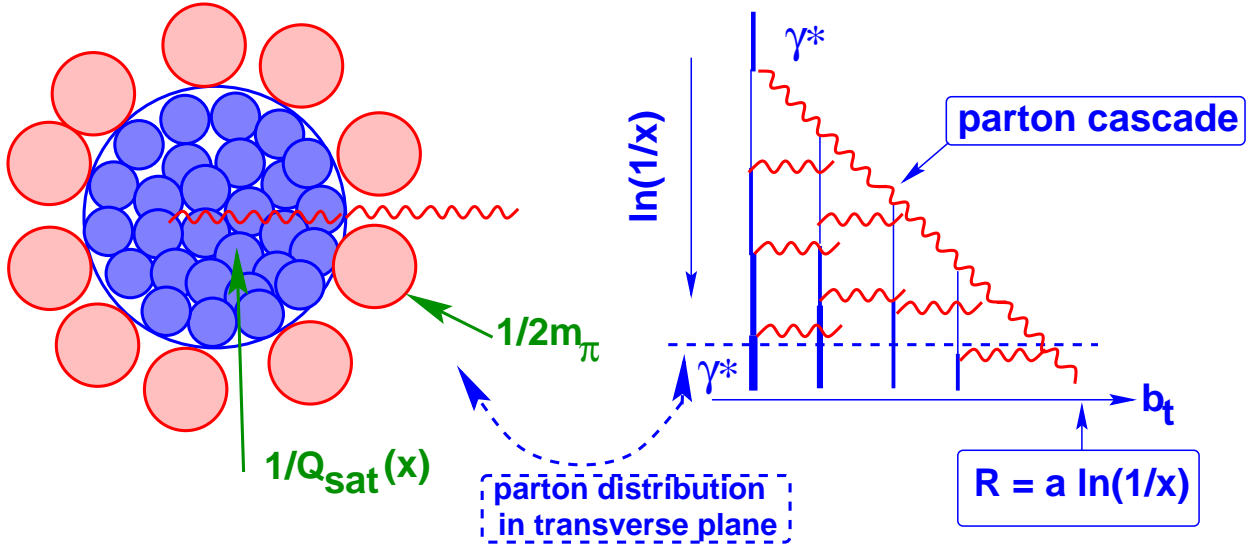,width=165mm}
\end{center}
\caption{ The structure of the parton cascade of the fast photon in the
frame where the second photon is at  rest. The picture is the
three-dimensional one since the  thickness of the vertical line reflects 
the value of the transverse momentum. The thicker the line the larger
value of the parton transverse momentum. The left part of the picture
shows the parton distribution in the transverse plane.}
 \label{pcsd}
\end{figure}

$N(x,r_{1,t},r_{2,t},b_t) $ denotes  the parton density, consequently 
 the fact that 
$N(x,r_{1,t},r_{2,t},b_t)
\,\rightarrow\,1$ can be understood as the \, fact  that the  partons 
reach a
maximal density at low $x$. This phenomenon is called saturation
\cite{GLR,MUQI,MV}. Therefore, at low $x$ we have the parton distribution
in the transverse plane presented in \fig{pcsd}: the uniform distribution
of partons ( dipoles) with   sizes of the order of $1/Q_{sat}(x)$ in
the disc
of radius $R(x)$. If one of the dipole inside of the disc will
emit one extra parton this emitted parton will interact with others partons
and as a result of this interaction its transverse momentum will be of the
order of $Q_{sat}(x)$. It means that this emitted gluon will not change
its position in impact parameter space since due to uncertainty principle 
\beq \label{UNPR}
\Delta \,b_t \,p_t\,\,\approx \,\,1
\eeq
its $\Delta b_t \,\approx\,1/Q_{sat}(x)$.
However, for the parton at the edge of the disc the  situation is different
since the emitted parton in the direction outside of the disc can move
freely without any interaction. This parton changes the size of the disc
by its displacement in $b_t$, namely $\Delta b_t
\,\approx\,1/p_t\,\approx\,\,1/2\,m_\pi$. In this estimate we consider the
non-perturbative emission with $p_t \approx 2m_\pi$ because,  as have been
discussed,  a  non-perturbative emission is needed to provide the 
unitarization of  our process. Since the emission that leads to a growth of
the disc occurs in one direction ( the exterior  of the disc) it leads to $R
\,=\,< | \Delta b_t | > \,n$ where $n$ is the number of emission at
given $x$. Since the emission  takes place at the edge of the disc
where the parton density is rather small,  $N(x,r_{1,t},r_{2,t},b_t)$ is determined by the 
BFKL
dynamics only \cite{LRREV,FIIM}. In the BFKL approach \cite{BFKL}$n\, =\,
\omega_l
\,\ln((x_0/x)$ since $
N^{OPS}\,\propto\,e^{n}\,=\,e^{\omega_l\,\ln(x_0/x)}$.
Therefore, we obtain \eq{SOL1}, namely,  $R(x)\,\equiv \,b_0(x)\,\,=
\,\,\frac{\omega_L}{2\,m_\pi}\,\ln(x_0/x)$.

We have discussed in this paper  the structure of dipole-dipole
interaction in the Glauber-Mueller approach which is the only one 
 on the market for the interaction of two dipoles of  the same sizes.
However, for two dipole with small but different sizes the non-linear
evolution equation \cite{GLR,MUQI,MV,KV} should be solved to which the
BFKL emission is only an approximation in the region of small partonic
densities. Comparison of the result of this paper with the dipole-dipole
interaction in , so called, double log approximation \cite{KL} shows that
the BFKL dynamics does not change   physics at large $b_t$. 
The non-linear evolution equation at fixed $b_t$ was solved \cite{LT} 
in the case when the BFKL kernel was replaced by the double log one. The
solution leads to the answer in the saturation region with geometrical
scale\cite{LT,GESC,FIIM}
\beq \label{GESCALE}
N(x,r_{1,t},r_{2,t};b_t)\,\,\,=
F\left(\tau\,=\,\,r^2_{1,t}\,\,Q^2_{sat}(x)\,\,e^{-\,4\,m_\pi\,b_t}\,\right)\,
\,.
\eeq
 
Therefore,  we believe that for the BFKL dynamics \eq{GESCALE} will hold.
This  belief is based on the similarity between double log and BFKL
approximation for $\gamma^* - \gamma^*$ processes.

{\bf  Acknowledgments:} 
  
We wish to thank  Jochen Bartels, Errol Gotsman and Uri Maor   for very
fruitful discussions on the subject.

One of us (E.L.)  would like to thank the DESY
Theory
Group for their hospitality and creative atmosphere during several
stages of this work. He is indebted to the Alexander-von-Humboldt Foundation for the award 
that gave him a possibility to work on low $x$ physics during the last year.

This research was supported in part  by the GIF grant $\#$ I-620-22.14/1999
  and by
Israeli Science Foundation, founded by the Israeli Academy of Science
and Humanities.

\begin{appendix}
\section{Appendix}
\setcounter{equation}{0} 
The integration over $R_1$ in \eq{BFKL1} can be taken explicitly 
\cite{NP,LI} and \eq{BFKL1} can be reduced to
\beq \label{A1} 
\int \,\,d^2\,R_1\,V(r_{1,t},R_1; \nu)\,\, V(r_{1,t},|\vec{R}_1 
\,-\,\vec{b}_t|; -\,\nu)\,\,=\,\,\frac{\nu^2}{( \frac{1}{4} \,+\,\nu^2 
)^2}  
\eeq
$$ c_1\,x^h\,{x^*}^h\,F(h,h,2h,x)\,F(h,h,2h,x^*) \,\,+\,\, c_2 x^{1 - 
h}{x^*}^{1-h}
F(1 - h,1 -h,2 - 2h,x)\,F(1 - h,1 -h,2 -2h,x^*) $$

where $F(\alpha,\beta,\gamma,z)$ is the hypergeometrical function (see 
Ref. \cite{RY} ); $x$ is the complex anharmonic ratio:
\beq \label{A2}
x\,\,=\,\,\frac{r_{1,t}\,r_{2,t}}{(b - z_1\,r_{1,t} -\bar{z}_2\,r_{2,t})
\,(b - \bar{z}_1\,r_{1,t} - z_2\,r_{2,t})}
\eeq
and $h\,=\,\frac{1}{2} \,+\,i\,\nu$. $x\,x^*$ gives 
\beq \label{A5}
x\,x^*\,\,=\,\,\frac{r^2_{1,t}\,r^2_{2,t}}{(\vec{b} - z_1\,\vec{r}_{1,t} 
-\bar{z}_2\,\vec{r}_{2,t})^2
\,(\vec{b} - \bar{z}_1\,\vec{r}_{1,t} - z_2\,\vec{r}_{2,t})^2}
\eeq
 One sees that \eq{A5} is invariant with respect to rotation in the plane.

The coefficients $c_1$ and $c_2$ have been calculated in Ref.\cite{LI} and 
they are equal:
\begin{eqnarray}
c_2 &=& \pi 2^{-1 + 4i\nu}\,\,
\frac{\Gamma( i\nu)}{\Gamma(\frac{1}{2} + i\nu)}\,; \label{A3}\\
\frac{c_1}{c_2} &=& - \left(\frac{\Gamma(2 - 
2h)}{\Gamma(2h)}\right)^2\,\left(\frac{\Gamma(h)}{\Gamma(1 - 
h)}\right)^4\,; \label{A4}
\end{eqnarray}

However, one can see that \eq{A1} does not reproduce the Born term of 
\eq{BAXS} at $y = 0$. To understand why it is so we should consider the 
vertex $V(r_{1,t},R_1; \nu)$ in momentum representation ( see Ref. 
\cite{NP}), namely,
\beq \label{A6}
V(r_{1,t},Q; \nu)\,\,\,=\,\,\,\int \,d^2\,R_1\,e^{ i 
\frac{\vec{Q}\,\cdot\,\vec{b}}{2}}\,V(r_{1,t},R_1; \nu)\,\,.
\eeq
It turns out \cite{NP} that
\begin{eqnarray}
V(r_{1,t},Q; \nu)\,\,\,&=&\,\,\,(Q\,Q^*)^{i\,\nu}2^{- 6i \nu}\,\Gamma^2(1 
- 
i 
\nu)\,\times\, \label{A7} \\
 & & \left(J_{-i\nu}(\frac{Q^* r_{1,t}}{4})\,J_{-i\nu}(\frac{Q 
r^*_{1,t}}{4}) \,\,-\,\,J_{i\nu}(\frac{Q^* 
r_{1,t}}{4})J_{i\nu}(\frac{Qr^*_{1,t}}{4})\,\right) \nonumber
\end{eqnarray}

At small $Q$ \eq{A7} leads to the following behavior of vertex 
$V(r-{1,t},Q; \nu)$:
\beq \label{A8}
V(r-{1,t},Q; \nu)\,\,\,\rightarrow |_{ Q^2 \,\rightarrow\, 
0}\,\,\,\,(\frac{r^2_t}{16})^{-i \nu}\,\left(\,1\,\,-\,\,(Q^2)^{2 
i\nu}\,(\frac{r^2_t}{16})^{i 2 \nu}\,\right)
\eeq
As have been discussed the matching with the Born approximation occurs at  
$i \nu\,\rightarrow\,\,\frac{1}{2}$. In this limit 
\beq \label{A9}
 V(r_{1,t},Q; \nu) \,\rightarrow\,\frac{r_{1,t}}{4}\,\left(\,1\,\, - 
\,\,Q^2\,\frac{r^2_t}{16}\,\right)\,\,,
\eeq
which has correct analytical behavior. Actually this behavior dictates 
the choice of the coefficients $c_1$ and $c_2$ in \eq{A3} and \eq{A4}. 
 
\begin{figure}
\begin{center}
\epsfig{file=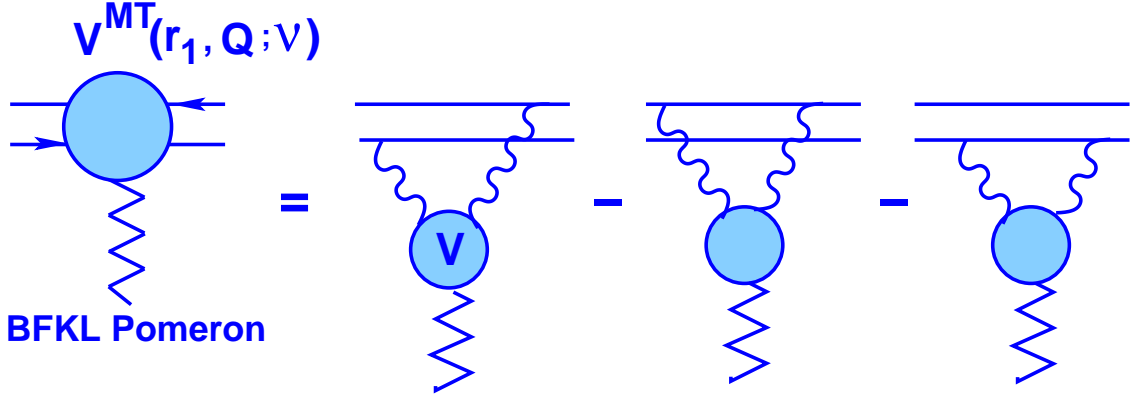,width=150mm}
\end{center}
\caption{Structure of the  Mueller-Tang vertex.}
\label{VT}
\end{figure}

However, at $ - i \nu\,\rightarrow\,\,\frac{1}{2}$ the low $Q$ behavior 
has a singularity $1/Q^2$. Therefore the symmetry of \eq{BFKL1} with 
respect to sign of $\nu$ is broken. Mueller and Tang \cite{MT} pointed out 
that this problem can be cured by adding to the expression of $ 
V(r_{1,t},Q; \nu) $ of \eq{V}, namely,
\begin{eqnarray} 
 V(r_{1,t},R_1; \nu) \,\,&\rightarrow& \,\, V^{MT}(r_{1,t},R_1; \nu)\,\,= 
\label{VMT} \\  
 & & V(r_{1,t},R_1; \nu) 
\,\,-\,\,\left(\frac{1}{(\vec{R}_1\,+\,\frac{1}{2}\,
\vec{r}_{1,t})^2}\right)^{\frac{1}{2}\,-\,i 
\nu}\,\,-\,\,\left(\frac{1}{(\vec{R}_1\,-\,\frac{1}{2}\,
\vec{r}_{1,t})^2}\right)^{\frac{1}{2}\,-\,i \,\nu}
 \nonumber
\end{eqnarray}
As was found\cite{LIP}  such terms can be added due to gauge invariance of 
QCD. In momentum representation (see \eq{A6}) $ V^{MT}(r_{1,t},Q; \nu)$ 
can be written as a sum of three terms as it is shown in \fig{VT}.

The Mueller-Tang vertex leads to the Born approximation amplitude in the 
form of \eq{BAXS}. However, as was discussed in Refs. 
\cite{LIP,BFLLRW,LI}, it has not been proven that this vertex will satisfy 
the BFKL equation. The solution in the form of \eq{A1} has a different 
form of the Born amplitude, namely, 
\beq \label{A10}
N^{BA} 
\,\propto\,\ln\left(\frac{\rho_{1,2}\,\rho_{1',2'}}{\rho_{1,2'}\,
\rho_{1',2}}\right)\,\ln 
\left(\frac{\rho_{1,2}\,\rho_{1',2'}}{\rho_{1,1'}\,
\rho_{2',2}}\right)\,\,. 
\eeq
However, these two expressions for the Born amplitude are equivalent due 
to gauge invariance of QCD \cite{LIP}.

Using \eq{A1} we can calculate  the dipole-dipole amplitude at $b_t =0$ 
and, therefore,  the saturation scale with better accuracy 
than within \eq{FINb0}. On the other hand the saturation momentum 
increases for $x \,\rightarrow\,0$ . Such an increase guarantees that 
\eq{FINb0} approaches the amplitude given by \eq{A1} in the region of low 
$x$. This is the reason why we prefer to use a simple solution of \eq{A1} 
instead of full expression of \eq{A1}.

It is easy to show that \eq{A1} describes all properties of diffusion 
approximation that has been discussed in section 2.

\end{appendix}

\end{document}